# The Influence of Fractional Derivatives on Thermodynamic Properties by Studying the CPSEHP Interaction

M. Abu-Shady [1*] and Sh. Y. Ezz-Alarab [1]

[1] Faculty of Science, Department of Mathematics and Computer Science, Menoufia University,

Shebin Elkom 32511, Egypt.

**Abstract**

The parametric Nikiforov-Uvarov (N-U) method is employed in conjunction with a generalized fractional derivative (GFD) to investigate the energy eigenvalues and the total normalized wave function associated with the Coulomb plus screened exponential hyperbolic potential (CPSEHP) in terms of Jacobi polynomials. This potential exhibits maximum effectiveness at lower values of the screening parameter. To explore the thermal and superstatistical characteristics, the derived energy eigenvalues are directly incorporated into the partition function (Z) and subsequently used to determine other thermodynamic quantities, including vibrational mean energy (U), specific heat capacity (C), entropy (S), and free energy (F). Comparisons with previous studies are conducted. The classical case is recovered from the fractional case by setting $\alpha = \beta = 1$, consistent with prior work. Our results demonstrate that the fractional parameter plays a crucial role in governing the thermal and superstatistical properties within the framework of this model.

## 1. Introduction

Over the last few decades, fractional calculus (FC) has attracted significant attention in a variety of scientific and engineering areas [1]. FC's prominence in several scientific and technical disciplines can be due to its benefits over numerical approaches, such as exact solutions and partial differential equations. Fractional differential equations are solved employing symmetry methods and perturbation analysis. In Ref. [2]. the radial Schrödinger equation (SE) is solved analytically. utilizing the conformable fractional variation of the Nikiforov-Uvarov (CF-NU) technique, the resulting dependent temperature potential in 3D and higher dimensions is used to determine the energy eigenvalues, correlated wave functions, and heavy quarkonium masses, such as charmonium and bottomonium, in a hot QCD medium. In Ref. [3], the conformable fractional derivative (CFD) method was used to investigate the fractional SE of a particle exhibiting position-dependent mass inside an infinity potential well. Considering any dependent thermal potential, trigonometric Rosen-Morse potential [4], hot-magnetized interactions potential [5], and generalized Cornell potential [6], The distinctive features for heavy quarkonium were examined using the N-dimensional radial SE within the framework of the CFD. Hammed et al. [7] used the CF-NU technique to generate triaxial nuclei solutions for the CF Bohr Hamiltonian using the Kratzer potential. Based on the Fermi-Pasta-Ulam model, the authors investigated the dependent upon time fraction fluctuation and the modified Gardner-type formula [8]. Abu-Shady and Kaabar proposed the generalized fractional derivative (GFD), which has more features than the earlier definitions [9]. Solving SE is a significant difficulty in quantum mechanics and particle physics for analyzing physical systems [10, 11, 12]. The application of the generalized fractional analytical iteration approach described in Ref. [13] towards the theory of single, double, and triple ground- state baryons is examined, and the hyper-central SE has been solved using this method.

Furthermore, as shown in Refs. [14, 15, 16], employing the NU technique allows for the exact solution of the SE and detailed system definition. This strategy outperforms previous methods, including those used in Ref. [17]. The Cornell potential and the expanded Cornell potential have already been used to examine the N-dimensional SE [24-28], applying wide methodologies including the NU method [22, 26, 29, 30]. Pekeris pattern prediction (PTA) [31, 32], the power series methodology (PST) [31], the asymptotic iteration procedure (AIM) [34], and the analysis of exact iteration approach (AEIM) [33].

The thermodynamic features of quantum structures are crucial in quantum physics and the physical sciences. Analyzing characteristics such as entropy, specific heat capacity, mean energy, and free energy necessitates the partition function, which is independent upon temperature [17, 34, 35].

Numerous academics have recently become interested in using a range of quantum potential models to investigate the thermodynamic characteristics of systems. For instance, Edet et al. [36] investigated the thermal characteristics of the Deng-fan Eckart potential model using the Poisson summation approach. Ikot et al. evaluated the thermodynamic properties of diatomic molecules with general molecular potential [37]. Onate [38] discovered the vibrational partition function, mean energy, vibrational specific heat capacity, and mean free energy in bound state formulations of the SE with the second Psychol-Teller-like potential. Within that paper, a hyperbolic variant of the Psychol-Teller-like potential was expressed. Numerous academics are interested in the practical use of the energy eigenvalue of the SE to study the partition function, thermodynamic properties, and superstatistics. lately, Okon et al. [39] applied the NU technique to study the thermodynamic characteristics and boundary phase formulations of two diatomic systems (carbon (II) oxide and scandium fluoride) using the Mobius square and screened Kratzer potential. their findings agreed with those of semiclassical WKB and others. They used a close form formulation of the energy eigenvalue to extract the partition function and other thermodynamic properties. Additionally, Oyewumi et al. [40] investigated the rotational-vibrational energy eigenvalues for diatomic systems employing the Pekeris-type approximated performance to centrifugal term and approximation solutions to the SE using the shifting Deng-Fan potential model within the asymptotic iteration framework. Furthermore, Boumali and Hassanabadi [41] investigated the thermal characteristics of a two-dimensional Dirac oscillator in the presence of an external magnetic field and discovered relativistic spin $\frac{1}{2}$ fermions influenced by Dirac oscillator coupling and a constant magnetic field in both computational and noncommutative spaces. Also, thermal properties of hadrons are investigated in Ref. [43, 44].

The main aim of the current work to study the effect the fractional parameter on superstatistics and thermal properties using the generalized fractional parametric NU method which The results in Ref. [49] is special case from the present model at α = β =1.

The paper is organized as follows: In Section 1, the contributions of earlier works are presented. In Section 2, the generalized fractional derivative is introduced. In Section 3, the radial Schrodinger equation with parametric generalized differential NU method is introduced. In Sections 4 and 5, thermodynamic properties and superstatistics formulation are presented. In Sec. 6, results and discussion are explained. In Section 7, conclusion is written.

# 2. The Generalized Fractional Derivative.

The present study suggests the generalized fractional derivative (GFD), it is an alternative formula for a fractional derivative. Compared to other traditional Caputo and Riemann-Liouville fractional derivative definitions, the generalized fractional derivative has been proposed to offer greater benefits, such that the derivative of two functions, See Ref. [45] for a new approach to easily computing fractional differential formulas. For a function $Z:(0,\infty)\rightarrow R$, the generalized fractional derivative of order $0<\alpha\leq 1$ of $Z(t)$ at $t>0$ is defined as

$$D^{GFD}Z(t)=\lim_{\varepsilon\rightarrow 0}\frac{Z\left(t+\frac{\Gamma(\beta)}{\Gamma(\beta-\alpha+1)}\varepsilon t^{1-\alpha}\right)-Z(t)}{\varepsilon};\beta>-1,\beta\in R^{+} \tag{1}$$

The generalized fractional derivative has the following characteristics:

$$D^{\alpha}[Z(t)]=k_1 t^{1-\alpha}\grave{Z}(t). \tag{2}$$

$$D^{\alpha}[D^{\alpha}Z(t)]=k_1^2\left[(1-\alpha)t^{1-2\alpha}\grave{Z}_{nl}(t)+t^{2-2\alpha}Z''(t)\right]. \tag{3}$$

where,

$$k_1=\frac{\Gamma[\beta]}{\Gamma[\beta-\alpha+1]},\text{ with }0<\alpha\leq 1,0<\beta\leq 1 \tag{4}$$

$$D^{\alpha}D^{\beta}t^{m}=D^{\alpha+\beta}t^{m}\text{ for function derivative of }Z(t)=t^{m},m\in R^{+}. \tag{5}$$

$$D^{GFD}(XY) = XD^{GFD}(\Upsilon) + \Upsilon D^{GFD}(X) \text{ where } X, \Upsilon \text{ be } \alpha - \text{ differentiable function} \tag{6}$$

$$D^{GFD}\left(\frac{X}{Y}\right) = \frac{\Upsilon D^{GFD}(X) - XD^{GFD}(\Upsilon)}{\Upsilon^2} \text{ where } X, \Upsilon \text{ be } \alpha \text{ - differentiable function} \tag{7}$$

$$D^{\alpha} I_{\alpha} Z(t) = Z(t) \text{ for } \geq 0 \text{ and Z is any function within the domain that is continuous} \tag{8}$$

## 2.1. The generalized fractional derivative with NU method

The parametric generalized fractional Nikiforov-Uvarov (NU) approach is introduced making use of generalized fractional derivative. In the fractional structure as in Ref [46], the second-order parametric generalized differential calculus is precisely computed.

$$D^{\alpha}[D^{\alpha}\psi(s)] + \frac{\bar{\tau}(s)}{\sigma(s)} D^{\alpha}\psi(s) + \frac{\bar{\sigma}(s)}{\sigma^2}\psi(s) = 0 \tag{9}$$

where, $\bar{\sigma}(s), \sigma(s)$ and $\bar{\tau}(s)$ are polynomials of $2\alpha$-th, $2\alpha$-th and $\alpha$-th degree.

$$\pi(s) = \frac{D^{\alpha}\sigma(s) - \bar{\tau}(s)}{2} \pm \sqrt{\left(\frac{D^{\alpha}\sigma(s) - \bar{\tau}(s)}{2}\right)^2 - \bar{\sigma}(s) + K\sigma(s)} \tag{10}$$

And,

$$\lambda = K + D^{\alpha}\pi(s) \tag{11}$$

$\pi(s)$ is $\alpha$-th degree polynomial, $\lambda$ is constant. It is feasible to identify the quantities of $K$ in the squareroot of Eq. (10), and therefore the formula within the square root is quadratic of expression. K is substituted for in Eq. (10), and we define.

$$\tau(s) = \bar{\tau}(s) + 2\pi(s) \tag{12}$$

Given that $\rho(s) > 0$ and $\sigma(s) > 0$, the derivative of $\tau$ is supposed to be negative in Ref. [47]. If $\lambda$ in Eq. (11) be

$$\lambda = \lambda_n = -nD^{\alpha}\tau - \frac{n(n-1)}{2} D^{\alpha}[D^{\alpha}\sigma(s)] \tag{13}$$

The solution of Eq. (3) is a combination of two independent components, and the hypergeometric type of equation has a distinctive solution with degree $\alpha$.

$$\psi(s) = \phi(s) y(s) \tag{14}$$

wherein,

$$y_n(s) = \frac{B_n}{\rho(s)} (D^{\alpha})^n (\sigma(s)^n \rho_n(s)) \tag{15}$$

$$D^{\alpha}[\sigma(s)\rho(s)] = \tau(s)\sigma(s), \tag{16}$$

$$\frac{D^{\alpha}\phi(s)}{\phi(s)} = \frac{\pi(s)}{\sigma(s)} \tag{17}$$

## 2.2. Parametric Second Order Generalized Differential Equation

The fractional Schrödinger formula is expressed in a general form as in Ref. [42] that follows.

$$D^{\alpha}[D^{\alpha}\psi(s)] + \frac{\bar{\tau}(s)}{\sigma(s)} D^{\alpha}\psi(s) + \frac{\bar{\sigma}}{(\sigma(s))^2}\psi(s) = 0 \tag{18}$$

where,

$$\bar{\tau}(s) = \alpha_1 - \alpha_2 s^{\alpha}, \tag{19}$$

$$\sigma(s) = s^{\alpha}(1 - \alpha_3 s^{\alpha}), \tag{20}$$

$$\bar{\sigma}(s) = -\xi_1 s^{2\alpha} + \xi_2 s^{\alpha} - \xi_3. \tag{21}$$

Substituting Eqs. (19), (20), and (21) into Eq. (10), we obtain.

$$\pi = \alpha_4 + \alpha_5 s^{\alpha} \pm \sqrt{(\alpha_6 - K_x\alpha_3)s^{2\alpha} + (\alpha_7 + K_x)s^{\alpha} + \alpha_8} \tag{22}$$

where,

$$\alpha_4 = \frac{1}{2}(k_1\alpha - \alpha_1), \tag{23}$$

$$\alpha_5 = \frac{1}{2}(\alpha_2 - 2\alpha_3 k_1\alpha), \tag{24}$$

$$\alpha_6 = \alpha_5^2 + \xi_1, \tag{25}$$

$$\alpha_7 = 2\alpha_4\alpha_5 - \xi_2, \tag{26}$$

$$\alpha_8 = \alpha_4^2 + \xi_3, \tag{27}$$

According to the NU technique, the square of a polynomial must be the function under square root in Eq. (22), which implies.

$$K_x = -(\alpha_7 + 2\alpha_3\alpha_8) \pm 2\sqrt{\alpha_8\alpha_9} \tag{28}$$

where,

$$\alpha_9 = \alpha_3\alpha_7 + \alpha_3^2\alpha_8 + \alpha_6 \tag{29}$$

In case $K_x$ is negative then has the form

$$K_x = -(\alpha_7 + 2\alpha_3\alpha_8) - 2\sqrt{\alpha_8\alpha_9} \tag{30}$$

So that $\pi$ becomes.

$$\pi = \alpha_4 + \alpha_5 s^\alpha - \left[\left(\sqrt{\alpha_9} + \alpha_3\sqrt{\alpha_8}\right)s^\alpha - \sqrt{\alpha_8}\right] \tag{31}$$

By using Eqs. (12), (22), and (31) then, we obtain.

$$\tau = \alpha_1 + 2\alpha_4 - (\alpha_2 - 2\alpha_5)s^\alpha - \left[\left(\sqrt{\alpha_9} + \alpha_3\sqrt{\alpha_8}\right)s^\alpha - \sqrt{\alpha_8}\right] \tag{32}$$

$$\begin{aligned} D^\alpha \tau &= k_1\left[-\alpha(\alpha_2 - 2\alpha_5) - 2\alpha\left(\sqrt{\alpha_9} + \alpha_3\sqrt{\alpha_8}\right)\right] \\ &= k_1\left[-2\alpha^2\alpha_3 - 2\alpha\left(\sqrt{\alpha_9} + \alpha_3\sqrt{\alpha_8}\right)\right] < 0 \end{aligned} \tag{33}$$

From Eqs. (2), and (32), we get.

We construct the energy spectrum equation from Eqs. (11), and (13)

$$\begin{aligned} &k_1\alpha\alpha_2 - (2n+1)k_1\alpha\alpha_5 + (2n+1)k_1\alpha\left(\sqrt{\alpha_9} + \alpha_3\sqrt{\alpha_8}\right) + n(n-1)k_1^2\alpha^2\alpha_3 \\ &\quad + \alpha_7 + 2\alpha_3\alpha_8 + 2\sqrt{\alpha_8\alpha_9} = 0. \end{aligned} \tag{34}$$

We obtain the standard formula of the energy eigenvalue as Ref [48], If $\alpha = \beta = 1$ then $k_1 = 1$,

$$\begin{aligned} &n\alpha_2 - (2n+1)\alpha_5 + (2n+1)\left(\sqrt{\alpha_9} + \alpha_3\sqrt{\alpha_8}\right) + n(n-1)\alpha_3 + \alpha_7 + 2\alpha_3\alpha_8 \\ &\quad + 2\sqrt{\alpha_8\alpha_9} = 0 \end{aligned} \tag{35}$$

by using Eq. (16), we get

$$\rho(s) = s^{\frac{\alpha_{10}-\alpha}{k_1}} (1 - \alpha_3 s^\alpha)^{\frac{\alpha_{11}}{\alpha_1\alpha_3} - \frac{\alpha_{10}}{\alpha k_1} - \frac{1}{k_1}}. \tag{36}$$

From Eq. (15), we obtain.

$$y_n = P_n^{\left(\frac{\alpha_{10}-\alpha}{k_1}, \frac{\alpha_{11}}{\alpha k_1\alpha_3} - \frac{\alpha_{10}}{\alpha k_1} - \frac{1}{k_1}\right)}(1 - 2\alpha_3 s^\alpha). \tag{37}$$

where, $L_n$ being the Laguerre polynomials, and K becomes.

$$\alpha_{10} = \alpha_1 + 2\alpha_4 + 2\sqrt{\alpha_8}$$

$$\alpha_{11} = \alpha_2 - 2\alpha_5 + 2\left(\sqrt{\alpha_9} + \alpha_3\sqrt{\alpha_8}\right) \tag{38}$$

The fractional wave function is given by Eq. (14),

$$\psi(s) = s^{\frac{\alpha_{12}}{k_1}} (1 - \alpha_3 s^\alpha)^{\frac{-\alpha_{13}}{\alpha k_1\alpha_3} - \frac{\alpha_{12}}{\alpha k_1}} P_n^{\left(\frac{\alpha_{10}-\alpha}{k_1}, \frac{\alpha_{11}}{\alpha k_1\alpha_3} - \frac{\alpha_{10}}{\alpha k_1} - \frac{1}{k_1}\right)}(1 - 2\alpha_3 s^\alpha) \tag{39}$$

where, $P_n^{(\gamma,\delta)}$ are Jacobi polynomials and

$$\alpha_{12} = \alpha_4 + \sqrt{\alpha_8}$$

$$\alpha_{13} = \alpha_5 - \left(\sqrt{\alpha_9} + \alpha_3\sqrt{\alpha_8}\right). \tag{40}$$

Some problems, in case $\alpha_3 = 0$.

$$\lim_{\alpha_3 \to 0} P_n^{\left(\frac{\alpha_{10}-\alpha}{k_1}, \frac{\alpha_{11}}{\alpha k_1 \alpha_3} - \frac{\alpha_{10}}{\alpha k_1} - \frac{1}{k_1}\right)}(1 - \alpha_3 s^{\alpha}) = L_n^{\frac{\alpha_{10}-\alpha}{k_1}}\left(\frac{\alpha_{11}}{\alpha k_1} s^{\alpha}\right), \tag{41}$$

$$\lim_{\alpha_3 \to 0} (1 - \alpha_3 s^{\alpha})^{\frac{-\alpha_{13}}{\alpha k_1 \alpha_3} - \frac{\alpha_{12}}{\alpha k_1}} = e^{\frac{\alpha_{13}}{\alpha k_1} s^{\alpha}} \tag{42}$$

Then Eq. (39), becomes.

$$\psi(s) = s^{\frac{\alpha_{12}}{k_1}} e^{\frac{\alpha_{13}}{\alpha k_1} s^{\alpha}} L_n^{\frac{\alpha_{10}-\alpha}{k_1}}\left(\frac{\alpha_{11}}{\alpha k_1} s^{\alpha}\right) \tag{43}$$

where, $L_n$ being the Laguerre polynomials, and K becomes.

$$K_x = -(\alpha_7 + 2\alpha_3\alpha_8) + 2\sqrt{\alpha_8\alpha_9} \tag{44}$$

then, the wave function is,

$$\psi(s) = s^{\frac{\alpha_{12}*}{k_1}} (1 - \alpha_3 s^{\alpha})^{\frac{-\alpha_{13}*}{\alpha_1\alpha_3} - \frac{\alpha_{12}*}{\alpha k_1}} P_n^{\left(\frac{\alpha_{10}*-\alpha}{k_1}, \frac{\alpha_{11}*}{\alpha k_1 \alpha_3} - \frac{\alpha_{10}*}{\alpha k_1} - \frac{1}{k_1}\right) \times}(1 - 2\alpha_3 s^{\alpha}), \tag{45}$$

The fractional energy eigenvalue formula is given:

$$\begin{aligned} &nk_1\alpha\alpha_2 - 2nk_1\alpha\alpha_5 + (2n+1)k_1\alpha\left(\sqrt{\alpha_9} - \alpha_3\sqrt{\alpha_8}\right) + n(n-1){k_1}^2\alpha^2\alpha_3 \\ &+\alpha_7 + 2\alpha_3\alpha_8 - 2\sqrt{\alpha_8\alpha_9} + k_1\alpha\alpha_5 = 0 \end{aligned} \tag{46}$$

where,

$$\alpha_{10^*} = \alpha_1 + 2\alpha_4 - 2\sqrt{\alpha_8} \tag{47}$$

$$\alpha_{11^*} = \alpha_2 - 2\alpha_5 + 2\left(\sqrt{\alpha_9} - \alpha_3\sqrt{\alpha_8}\right) \tag{48}$$

$$\alpha_{12^*} = \alpha_4 - \sqrt{\alpha_8} \tag{49}$$

$$\alpha_{13}^* = \alpha_5 - \left(\sqrt{\alpha_9} - \alpha_3\sqrt{\alpha_8}\right), \tag{50}$$

# 3. The radial Schrodinger equation with Parametric Generalized Differential NU method

The suggested Coulomb plus screened hyperbolic exponential potential (CPSHEP) [49] is presented so as.

$$V(r) = -\frac{v_1}{r} + \left(\frac{B}{r} - \frac{v_2 \cosh \alpha_x}{r^2}\right) e^{-\alpha_x r} \tag{51}$$

where $\alpha_x$ is the adjustable screening parameter, $B$ is a real constant parameter, and $v_1$ and $v_2$ are the potential depths. This is the centrifugal term's Pekeris-like approximation:

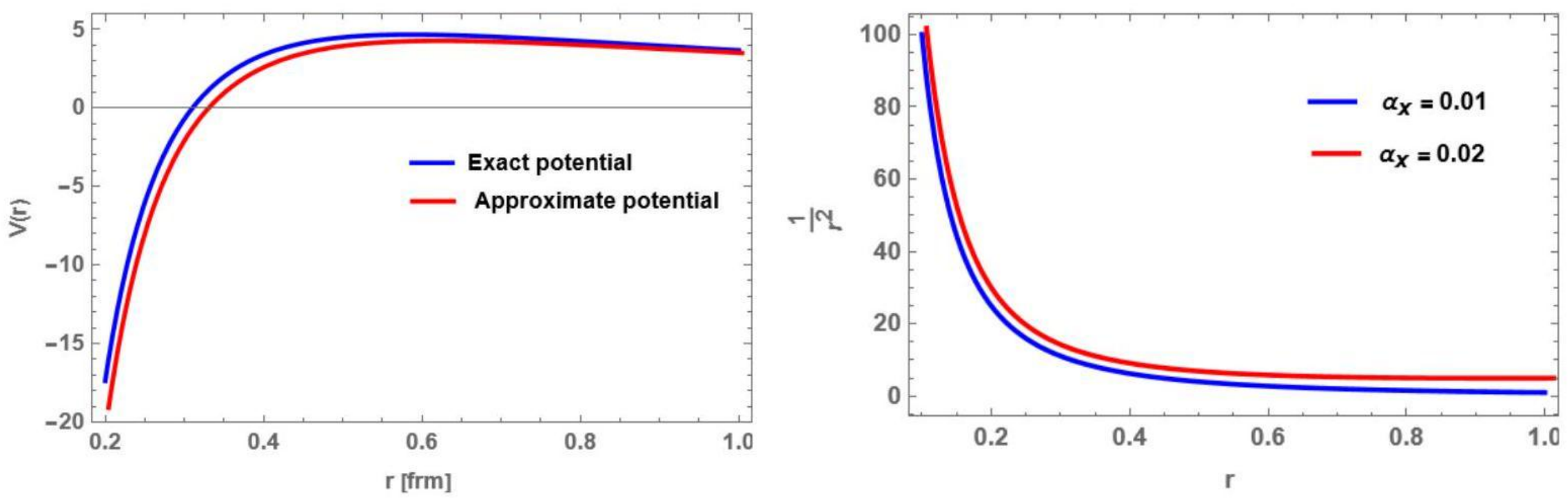

Figure 1: (the left panel), the potential $V(r)$ of the exact potential via approximation potential are drawn as a function of distance $(r)$ at $\alpha_x = 0.2$, (the right panel), the graph of Pekeris approximation for various values of $\alpha_x$.

$$\frac{1}{r^2} = \frac{\alpha_x^2}{(1 - e^{-\alpha_x r})^2} \Rightarrow \frac{1}{r} = \frac{\alpha_x}{(1 - e^{-\alpha_x r})}. \tag{52}$$

by substituting Eq. (52) to Eq. (51), then we obtain the approximation form of the mentioned potential as follows.

$$V(r) = -\frac{v_1 \alpha_x}{(1 - e^{-\alpha_x r})} + \left(\frac{B\alpha_x}{(1 - e^{-\alpha_x r})} - \frac{v_2\, \alpha_x^2 \cosh \alpha_x}{(1 - e^{-\alpha_x r})^2}\right) e^{-\alpha_x r} \tag{53}$$

(54)

The radial Schrödinger equation with the centrifugal term is given as follows.

$$\frac{d^2 R(r)}{dr^2} + \frac{2\,\mu}{\hbar^2} \cdot \left\{E - V(r) - \frac{\hbar^2\, l(l+1)}{2\,\mu\, r^2}\right\} R(r) = 0$$

The exact solution to Eq. (54) can only be obtained through analytical solution when angular orbital quantum number l = 0. Nevertheless, Eq. (54) can only be solved for l > 0 by applying the centrifugal term estimates in Eq. (52). Eq. (51) can be substituted into Eq. (54) to yield.

$$\frac{d^2R(r)}{dr^2}+\frac{2\,\mu}{\hbar^2}\cdot\left\{E+\frac{v_1}{\mathrm{r}}-\frac{\mathrm{B}\,e^{-\alpha_x r}}{\mathrm{r}}+\frac{v_2\,e^{-\alpha_x r}\cosh\alpha_x}{r^2}-\frac{\hbar^2\,l(l+1)}{2\,\mu\,r^2}\right\}R(r)=0 \qquad (55)$$

Eq. (52) can be substituted into Eq. (55) to obtain the following equation:

$$\frac{d^2R(r)}{dr^2}+\frac{2\,\mu}{\hbar^2}\cdot\left\{E+\frac{v_1\alpha_x}{(1-e^{-\alpha_x r})}-\frac{\mathrm{B}\,\alpha_x\;e^{-\alpha_x r}}{(1-e^{-\alpha_x r})}+\frac{v_2\,\alpha_x^2\,e^{-\alpha_x r}\cosh\alpha_x}{(1-e^{-\alpha_x r})^2}-\frac{\hbar^2\alpha_x^2\,l(l+1)}{2\,\mu\,(1-e^{-\alpha_x r})^2}\right\}R(r)=0 \qquad (56)$$

With a little mathematical simplification and a definition of s= $e^{-\alpha_x r}$, Eq. (56) will be expressed as follows.

$$\frac{d^2R(s)}{ds^2}+\frac{(1-s)}{s(1-s)}\frac{dR}{ds}+\frac{1}{s^2(1-s)^2}\cdot\left\{\begin{matrix}-(\varepsilon^2-\chi_1)s^2\\+(2\varepsilon^2-\delta^2-\chi_1+\chi_2)s\\-(\varepsilon^2-\delta^2+l(l+1))\end{matrix}\right\}R(s)=0 \qquad (56)$$

where,

$$\varepsilon^2=-\frac{2\mu E_{nl}}{\hbar^2\alpha_x^2},\delta^2=\frac{2\mu v_1}{\hbar^2\alpha_x},\chi_1=\frac{2\mu B}{\hbar^2\alpha_x},\chi_2=\frac{2\mu v_2\cosh\alpha_x}{\hbar^2},s=e^{-\alpha_x r}. \qquad (57)$$

Consequently, we obtain the generalized fractional radial constituent of the Schrödinger equation [50] as follows.

$$D^\alpha[D^\alpha R(s)]+\frac{1-s^\alpha}{s^\alpha(1-s^\alpha)}D^\alpha R(s)+\frac{-\xi_1 s^{2\alpha}+\xi_2 s^\alpha-\xi_3}{\left(s^\alpha(1-s^\alpha)\right)^2}R(s)=0 \qquad (58)$$

Comparing Eq. (57) to Eq. (59), the following polynomials were obtained:

$$\xi_1=(\varepsilon^2-\chi_1),\xi_2=(2\varepsilon^2-\delta^2-\chi_1+\chi_2),\xi_3=\varepsilon^2-\delta^2+l(l+1) \qquad (60)$$

Using Eqs. (10), (12), from Eq. (19) to Eq. (40), other parametric constants are obtained as follows:

$$\alpha_1=1,\alpha_2=1,\alpha_3=1,\alpha_4=\frac{1}{2}(k_1\alpha-1). \qquad (61)$$

$$\alpha_5=\frac{1}{2}(1-2k_1\alpha),\alpha_6=\frac{1}{4}(1-2k_1\alpha)^2-\frac{2\mu E_{nl}}{\hbar^2\alpha_x^2}-\frac{2\mu B}{\hbar^2\alpha_x}. \qquad (62)$$

$$\alpha_7=\frac{1}{2}(k_1\alpha-1)(1-2k_1\alpha)-2-\frac{2\mu E_{nl}}{\hbar^2\alpha_x^2}+\frac{2\mu v_1}{\hbar^2\alpha_x}+\frac{2\mu B}{\hbar^2\alpha_x}-\frac{2\mu v_2\cosh\alpha_x}{\hbar^2}. \qquad (63)$$

$$\alpha_8=\frac{1}{4}(k_1\alpha-1)^2-\frac{2\mu E_{nl}}{\hbar^2\alpha_x^2}-\frac{2\mu v_1}{\hbar^2\alpha_x},\alpha_9=\frac{1}{4}k_1^2\alpha^2-\frac{2\mu v_2\cosh\alpha_x}{\hbar^2}+l(l+1). \qquad (64)$$

$$\alpha_{10} = k_1\alpha + 2\sqrt{\frac{1}{4}(k_1\alpha - 1)^2 - \frac{2\mu E_{nl}}{\hbar^2\alpha_x^2} - \frac{2\mu v_1}{\hbar^2\alpha_x} + l(l+1)}. \tag{65}$$

$$\alpha_{11} = 2k_1\alpha + 2\sqrt{\frac{1}{4}k_1^2\alpha^2 - \frac{2\mu v_2\cosh\alpha_x}{\hbar^2} + l(l+1)} \tag{66}$$

$$+\sqrt{\frac{1}{4}(k_1\alpha - 1)^2 - \frac{2\mu E_{nl}}{\hbar^2\alpha_x^2} - \frac{2\mu v_1}{\hbar^2\alpha_x} + l(l+1)} \tag{67}$$

$$\alpha_{12} = \frac{1}{2}(k_1\alpha - 1) + \sqrt{\frac{1}{4}(k_1\alpha - 1)^2 - \frac{2\mu E_{nl}}{\hbar^2\alpha_x^2} - \frac{2\mu v_1}{\hbar^2\alpha_x} + l(l+1)} \tag{68}$$

$$\alpha_{13} = \frac{1}{2}(1 - 2k_1\alpha) - \frac{1}{2}\sqrt{k_1^2\alpha^2 - \frac{2\mu v_2\cosh\alpha_x}{\hbar^2} - l(l+1)} + \sqrt{\frac{1}{4}(k_1\alpha - 1)^2 - \frac{2\mu E_{nl}}{\hbar^2\alpha_x^2} - \frac{2\mu v_1}{\hbar^2\alpha_x} + l(l+1)} \tag{69}$$

The energy eigenvalue in the fractional form for the present potential can be determined by Eqs. (34), (60), and from Eq. (19) to Eq. (40) with considerable algebraic simplification as follows.

$$E_{nl} = \frac{\hbar^2\alpha_x^2 l(l+1)}{2\mu} - v_1\alpha_x + \frac{\hbar^2\alpha_x^2(k_1\alpha - 1)^2}{8\mu} + \frac{\hbar^2\alpha_x^2}{2\mu} * \cdot\left\{\frac{F_1 + F_2 + F_3}{k_1\alpha * (2n+1) + \sqrt{K_1^2\alpha^2 + 4l(l+1) - (8v_2\mu\cosh\alpha_x/\hbar^2)}}\right\}^2. \tag{70}$$

we can get the classical case from Eq. (70) at $\alpha = \beta = 1$ then $K_1 = 1$ [49], as the following equation.

$$E_{nl} = \frac{\hbar^2\alpha_x^2 l(l+1)}{2\mu} - v_1\alpha_x + \frac{\hbar^2\alpha_x^2}{2\mu} * \cdot\left\{\frac{F_4 + F_5 + F_6}{(2n+1) + \sqrt{1 + 4l(l+1) - \left(\frac{8v_2\mu\cosh\alpha_x}{\hbar^2}\right)}}\right\}^2. \tag{59}$$

where,

$$F_1 = K_1^2\alpha^2(n^2 + n + (1/2)). \tag{72}$$

$$F_2 = k_1\alpha * (n + (1/2))\sqrt{K_1^2\alpha^2 + 4l(l+1) - (8v_2\mu\cosh\alpha_x/\hbar^2)}. \tag{73}$$

$$F_3 = (2\mu B/\hbar^2\alpha_x) - \left(\frac{2\mu v_1}{\hbar^2\alpha_x}\right) - \left(\frac{2v_2\mu\cosh\alpha_x}{\hbar^2}\right) + 2l(l+1). \tag{74}$$

$$F_4 = (n^2 + n + (1/2)). \tag{75}$$

$$F_5 = \left(n + \left(\frac{1}{2}\right)\right)\sqrt{1 + 4l(l+1) - \left(\frac{8v_2\mu\cosh\alpha_x}{\hbar^2}\right)}. \tag{76}$$

$$F_6 = (2\mu B/\hbar^2\alpha_x) - \left(\frac{2\mu v_1}{\hbar^2\alpha_x}\right) - \left(\frac{2v_2\mu\cosh\alpha_x}{\hbar^2}\right) + 2l(l+1). \tag{77}$$

The generalized fractional of the total wave function using Eqs. [38-40], is given as

$$\Psi_{nl}(s) = N_{nl}s^{(F_7)} * (1 - s^{\alpha})^{(F_8+F_9)} * P_n^{\left(F_{10},F_{11}+F_{12}-\frac{1}{k_1}\right)} * (1 - 2s^{\alpha}). \tag{60}$$

$$F_7 = \frac{\frac{1}{2}(k_1\alpha - 1) + \sqrt{\frac{1}{4}(k_1\alpha - 1)^2 - \frac{2\mu E_{nl}}{\hbar^2\alpha_x^2} - \frac{2\mu v_1}{\hbar^2\alpha_x} + l(l+1)}}{k_1}. \tag{61}$$

$$F_8 = -\left(\frac{1}{2}(1 - 2k_1\alpha) - \frac{1}{2}\sqrt{k_1^2\alpha^2 - \frac{2\mu v_2\cosh\alpha_x}{\hbar^2} - l(l+1)} + \frac{\sqrt{\frac{1}{4}(k_1\alpha - 1)^2 - \frac{2\mu E_{nl}}{\hbar^2\alpha_x^2} - \frac{2\mu v_1}{\hbar^2\alpha_x} + l(l+1)}}{\alpha k_1}\right) \tag{80}$$

$$F_9 = -\frac{\frac{1}{2}(k_1\alpha - 1) + \sqrt{\frac{1}{4}(k_1\alpha - 1)^2 - \frac{2\mu E_{nl}}{\hbar^2\alpha_x^2} - \frac{2\mu v_1}{\hbar^2\alpha_x} + l(l+1)}}{\alpha k_1} \tag{81}$$

$$F_{10} = \frac{k_1\alpha + 2\sqrt{\frac{1}{4}(k_1\alpha - 1)^2 - \frac{2\mu E_{nl}}{\hbar^2\alpha_x^2} - \frac{2\mu v_1}{\hbar^2\alpha_x} + l(l+1)} - \alpha}{k_1} \tag{82}$$

$$F_{11} = 2k_1\alpha + 2\sqrt{\frac{1}{4}k_1^2\alpha^2 - \frac{2\mu v_2\cosh\alpha_x}{\hbar^2} + l(l+1)} + \sqrt{\frac{1}{4}(k_1\alpha - 1)^2 - \frac{2\mu E_{nl}}{\hbar^2\alpha_x^2} - \frac{2\mu v_1}{\hbar^2\alpha_x} + l(l+1)} \tag{62}$$

$$F_{12} = -\frac{k_1\alpha + 2\sqrt{\frac{1}{4}(k_1\alpha - 1)^2 - \frac{2\mu E_n l}{\hbar^2 \alpha_x^2} - \frac{2\mu v_1}{\hbar^2 \alpha_x} + l(l+1)}}{\alpha k_1} \quad (63)$$

The special case of the total un-normalized wave function at $\alpha = \beta = 1$ then $K_1 = 1$, as in agreement with [49] as following.

$$\Psi_{nl}(s) = N_{nl} s^{\sqrt{(\varepsilon^2-\delta^2+l(l+1))}} (1s)^{-\frac{1}{2}-2\sqrt{(\varepsilon^2-\delta^2+l(l+1))}-\frac{1}{2}\sqrt{1+4l(l+1)-4\chi_2}} P_n^{\left[\left(2\sqrt{(\varepsilon^2-\delta^2+l(l+1))}\right)\cdot\left(\sqrt{(1+4l(l+1)-}\right.\right.} - 2s) \quad (85)$$

# 4. Thermodynamic Properties

In this section, we describe the potential model's thermodynamic properties. The precise partition function offered can be used to determine the thermodynamic characteristics of quantum systems as follows.

$$Z(\beta_1) = \sum_{n=0}^{\lambda} e^{-\beta_1 E_n} \quad (86)$$

where $\lambda$ is an upper constraint on the vibrational quantum number determined by the numerical solution of $dE_n/dn = 0$, expressed as $\lambda = \frac{1}{k_1\alpha}\left(-\delta + \sqrt{\delta(k_1\alpha - \delta) + Q_3}\right)$, $\beta_1 = 1/kT$, where $k$ and $T$ are the Boltzmann constant and absolute temperature, correspondingly. The integral in Eq. (87) may substitute the place of the summation in the classical limit.

$$Z(\beta_1) = \int_0^{\lambda} e^{-\beta_1 E_n} dn \quad (87)$$

The energy eigen value equation as in Eq. (70) can be expressed in a short and concise manner as follows to yield the partition functions.

$$E_{nl} = Q_1 + Q_2 * \left\{\frac{(k_1\alpha n + \delta)}{2} + \frac{\delta(k_1\alpha - \delta) + Q_3}{2(k_1\alpha n + \delta)}\right\}^2, \quad (88)$$

where,

$$
Q_1 = \frac{\hbar^2 \alpha_x^2 l(l+1)}{2\mu} - v_1 \alpha + \frac{\hbar^2 \alpha_x^2 (k_1 \alpha - 1)^2}{8\mu}, Q_2 = \frac{\hbar^2 \alpha_x^2}{2\mu}.
$$

$$
Q_3 = \left(\frac{2\mu B}{\hbar^2 \alpha_x}\right) - \left(\frac{2\mu v_1}{\hbar^2 \alpha_x}\right) - \left(\frac{2 v_2 \mu \cosh \alpha_x}{\hbar^2}\right) + 2l(l+1)
$$

$$
\delta = \frac{1}{2} k_1 \alpha + \sqrt{\frac{1}{4} K_1^2 \alpha^2 - \frac{2 v_2 \mu \cosh \alpha_x}{\hbar^2} + l(l+1).} \tag{89}
$$

the following form can be used to represent Eq. (88) as follows.

$$
E_{nl} = (Q_1 + \frac{Q_2(\delta(k_1\alpha - \delta) + Q_3)}{2}) + (\frac{Q_2 \rho^2}{4} + \frac{Q_2(\delta(k_1\alpha - \delta) + Q_3)^2}{4\rho^2}) \tag{90}
$$

where,

$$
\rho = k_1 \alpha n + \delta \tag{91}
$$

**(i)** Partition function is obtained by inserting equation Eq. (90) to Eq. (87) and noting modifications to the integration boundaries utilizing equation Eq. (91)

$$
= e^{-\beta_1\left(Q_1 + \frac{Q_2(\delta(k_1\alpha-\delta)+Q_3)}{2}\right)} \int_{\delta}^{k_1\alpha\lambda+\delta} e^{\beta_1(\frac{Q_2\rho^2}{4} + \frac{Q_2(\delta(k_1\alpha-\delta)+Q_3)^2}{4\rho^2})} d\rho \tag{64}
$$

The partition function of Eq. (92) is obtained as

$$
Z = \frac{(\sqrt{\pi} e^{-\frac{1}{2}\beta_1(NQ_2 + NN_0^2 + 2Q_1)} (\operatorname{erf}(\sqrt{\beta_1}\Lambda_1) + \operatorname{erf}(\sqrt{\beta_1}\Lambda_3)) - e^{\beta_1 NN_0^2} \operatorname{erf}(\sqrt{\beta_1}\Lambda_2) + e^{\beta_1 NN_0^2} \operatorname{erf}(\sqrt{\beta_1}\Lambda_4))}{2\sqrt{\beta_1} N_0} \tag{65}
$$

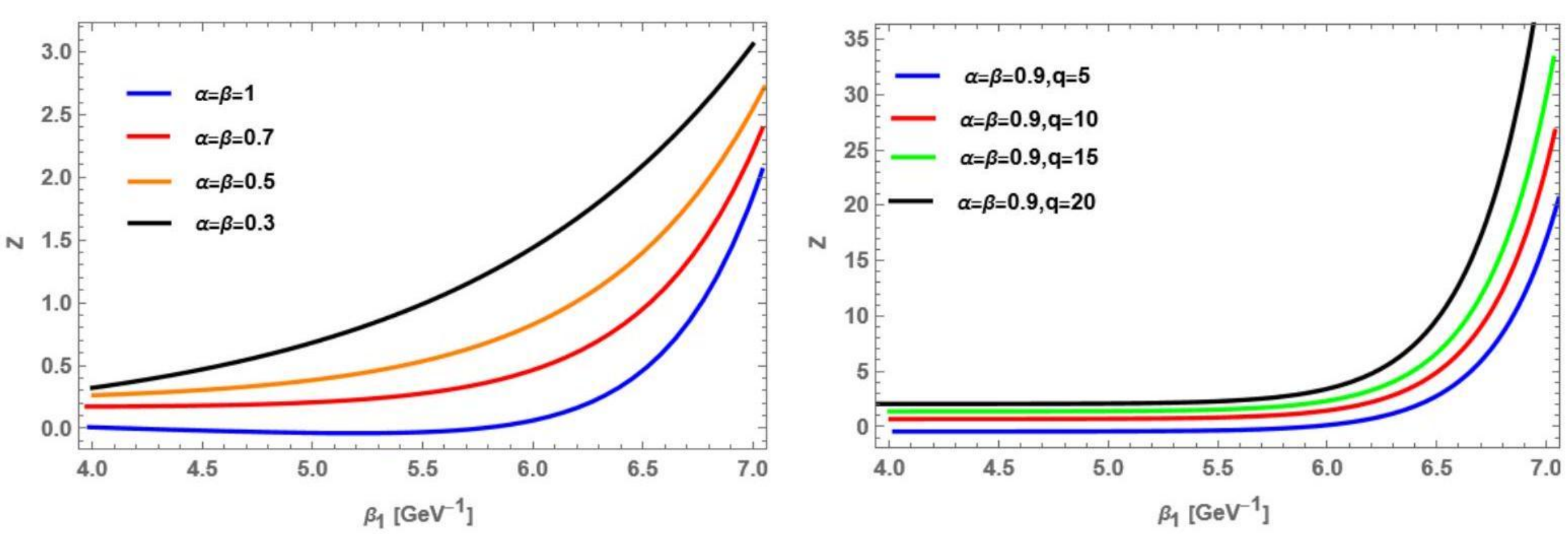

Figure 2: (The left panel), the partition function (Z) is displayed as a function of $\beta_1$, for various values of $\alpha$, and $\beta$. (The right panel), the partition function (Z) is displayed as a function of $\beta_1$, for various values of deformed parameter q for CPSHEP at $\alpha = \beta = 0.9$

**(ii)** Vibrational free energy is given as follows.

$$F(\beta_1) = -\frac{1}{\beta_1}\, lnZ(\beta_1). \tag{66}$$

$$F = -\frac{\log\left(\frac{\sqrt{\pi}e^{-\frac{1}{2}\beta_1\left(NQ_2+NN_0^2+2Q_1\right)}\left(\mathrm{erf}(\sqrt{\beta_1}\Lambda_1)+\mathrm{erf}(\sqrt{\beta_1}\Lambda_3)-e^{\beta_1 NN_0^2 \mathrm{erf}(\sqrt{\beta_1}\Lambda_2)+e^{\beta_1 NN_0^2}\mathrm{erf}(\sqrt{\beta_1}\Lambda_4)}\right)}{2\sqrt{\beta_1}N_0}\right)}{\beta_1} \tag{95}$$

**(iii)** Vibrational mean energy is given as follows.

$$U(\beta_1) = -\frac{d}{d\beta_1} lnZ(\beta_1). \tag{96}$$

$$U = \frac{1}{2}\left(\frac{1}{\beta_1} - \Omega + NQ_2 + NN_0^2 + 2Q_1\right), \tag{97}$$

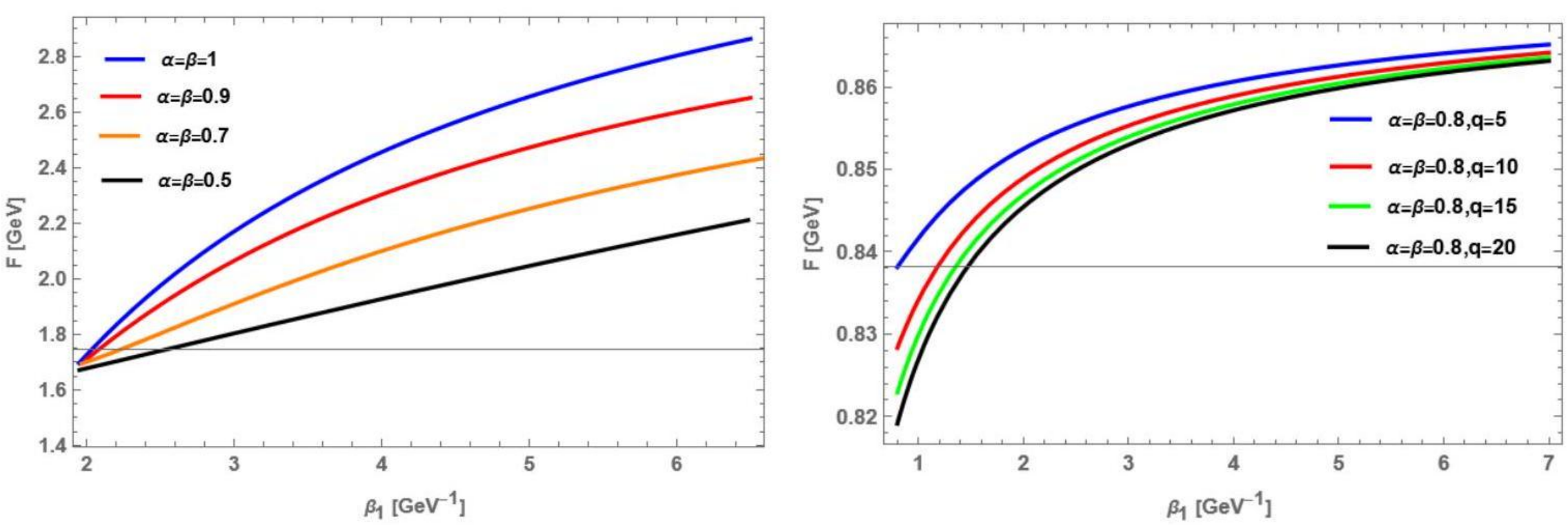

Figure 3: (The left panel), the free energy (F) is displayed for various values of $\alpha$ as a function of $\beta_1$, and $\beta$. (The right panel), F is displayed as a function of $\beta_1$, for various values of deformed parameter q for CPSHEP at $\alpha = \beta = 0.8$.

$$\Omega = \frac{2(F_{13} - F_{14})}{\sqrt{\pi}\sqrt{\beta_1}\left(\mathrm{erf}(\sqrt{\beta_1}\Lambda_1) + \mathrm{erf}(\sqrt{\beta_1}\Lambda_3) - e^{\beta_1 NN_0^2}\mathrm{erf}(\sqrt{\beta_1}\Lambda_2) + e^{\beta_1 NN_0^2 \mathrm{erf}(\sqrt{\beta_1}\Lambda_4)}\right)}, \tag{98}$$

$$F_{13} = \Lambda_1 e^{-\beta_1\Lambda_1^2} + \Lambda_3 e^{-\beta_1\Lambda_3^2} + \sqrt{\pi}\sqrt{\beta_1}NN_0^2\left(-e^{\beta_1 NN_0^2}\right)\left(\mathrm{erf}(\sqrt{\beta_1}\Lambda_2) - \mathrm{erf}(\sqrt{\beta_1}\Lambda_4)\right) \tag{99}$$

$$F_{14} = -\Lambda_2 e^{\beta_1\left(NN_0^2-\Lambda_2^2\right)} + \Lambda_4 e^{\beta_1\left(NN_0^2-\Lambda_4^2\right)} \tag{100}$$

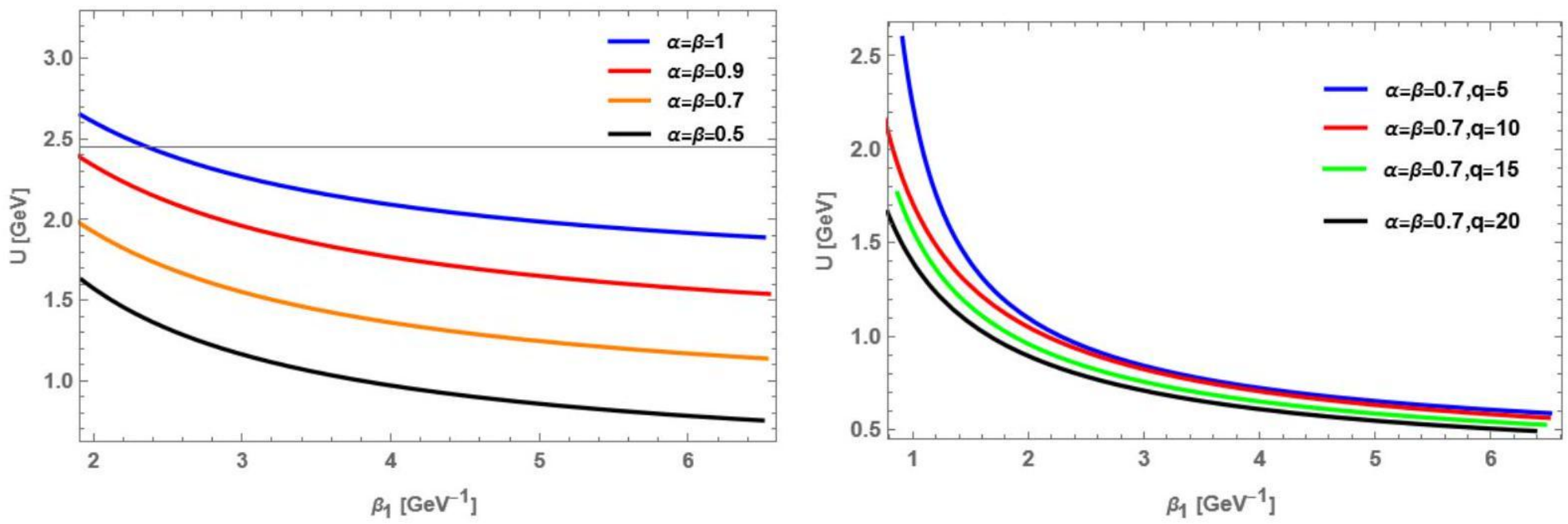

Figure 4: (The left panel), the mean energy (U) is displayed as a function of $\beta_1$ for various values of $\alpha$, and $\beta$. (The right panel), the mean energy is displayed as a function of $\beta_1$, for various values of deformed parameter q for CPSHEP at $\alpha = \beta = 0.7$.

**(iv)** Vibrational entropy is given as follows.

$$S(\beta_1) = K_1 \ ln\, Z(\beta_1) + K_1\, \beta_1 \frac{d}{d\beta_1} lnZ(\beta_1) \tag{101}$$

$$S = K_1\left(\frac{\sqrt{\beta_1}(F_{15}+F_{16})}{F_{17}} + \log(F_{18}) + \frac{1}{2}\beta_1(NQ_2 + NN_0^2 + 2Q_1) + \frac{1}{2}\right) \tag{102}$$

$$F_{15} = -\Lambda_1 e^{-\beta_1\Lambda_1^2} - \Lambda_3 e^{-\beta_1\Lambda_3^2} + \sqrt{\pi}\sqrt{\beta_1}NN_0^2 e^{\beta_1 NN_0^2}\left(\mathrm{erf}\left(\sqrt{\beta_1}\Lambda_2\right) - \mathrm{erf}\left(\sqrt{\beta_1}\Lambda_4\right)\right). \tag{103}$$

$$F_{16} = \Lambda_2 e^{\beta_1(NN_0^2-\Lambda_2^2)} - \Lambda_4 e^{\beta_1(NN_0^2-\Lambda_4^2)}. \tag{104}$$

$$F_{17} = \pi\left(\mathrm{erf}\left(\sqrt{\beta_1}\Lambda_1\right) + \mathrm{erf}\left(\sqrt{\beta_1}\Lambda_3\right) - e^{\beta_1 NN_0^2}\,\mathrm{erf}\left(\sqrt{\beta_1}\Lambda_2\right) + \ e^{\beta_1 NN_0^2}\mathrm{erf}\left(\sqrt{\beta_1}\Lambda_4\right)\right). \tag{105}$$

$$F_{18} = \frac{\sqrt{\pi}e^{-\frac{1}{2}\beta_1(NQ_2+NN_0^2+2Q_1)}\begin{pmatrix}\mathrm{erf}\left(\sqrt{\beta_1}\Lambda_1\right) + \mathrm{erf}\left(\sqrt{\beta_1}\Lambda_3\right)\\ -e^{\beta_1 NN_0^2}\mathrm{erf}\left(\sqrt{\beta_1}\Lambda_2\right) + e^{\beta_1 NN_0^2}\mathrm{erf}\left(\sqrt{\beta_1}\Lambda_4\right)\end{pmatrix}}{2\sqrt{\beta_1}N_0}. \tag{106}$$

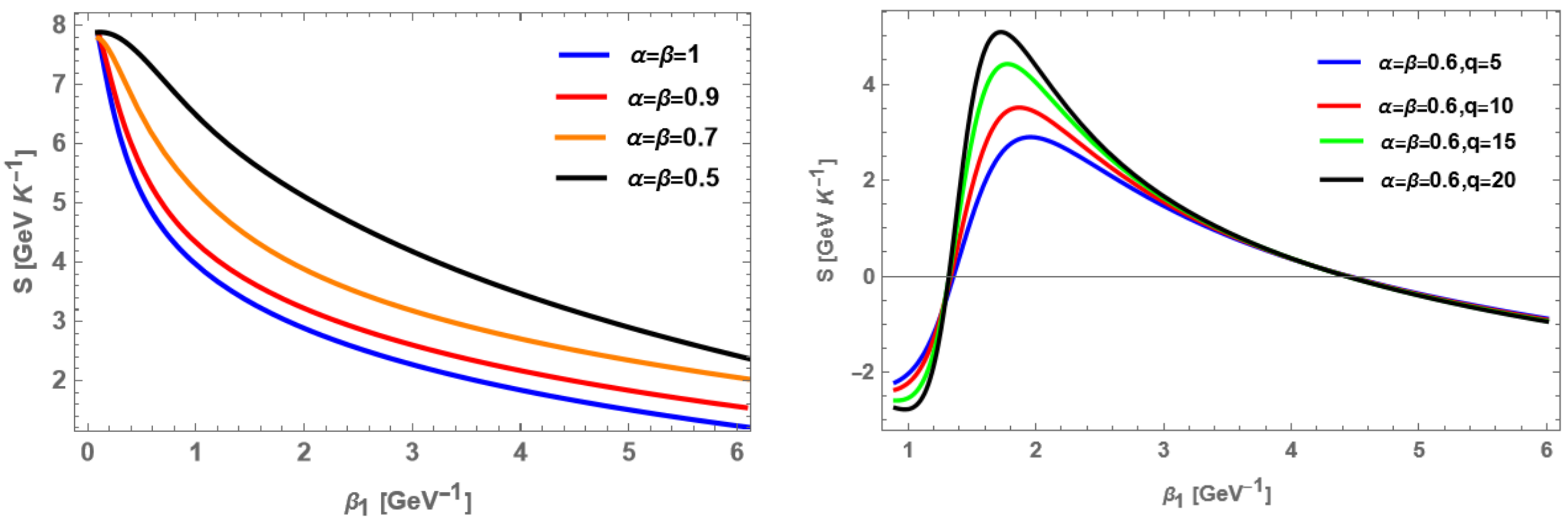

Figure 5: (The left panel), the entropy (S) is displayed as a function of $\beta_1$ for various values of $\alpha$, and $\beta$. (The right panel), S is displayed as a function of $\beta_1$, for various values of deformed parameter q for CPSHEP at $\alpha = \beta = 0.6$.

**(v)** Vibrational specific heat capacity is given as follows.

$$C(\beta_1) = -K_1 \beta_1^{\ 2} \frac{dU(\beta_1)}{d\beta_1}. \tag{107}$$

$$C = \frac{1}{2} K_1 \left( \frac{2\beta_1^{3/2}(F_{19} + F_{20})}{F_{21}} - \frac{\sqrt{\beta_1}(F_{22})}{F_{21}} - \frac{2e^{-2\beta_1(\Lambda_1^2+\Lambda_2^2+\Lambda_3^2+\Lambda_4^2)}\beta_1(F_{23})^2}{\pi(F_{21})^2} + 1 \right). \tag{108}$$

$$\begin{aligned} F_{19} = &-e^{NN_0^2\beta_1} N^2 \left(\mathrm{erf}\left(\sqrt{\beta_1}\Lambda_2\right) - \mathrm{erf}\left(\sqrt{\beta_1}\Lambda_4\right)\right)\sqrt{\pi}\sqrt{\beta_1} N_0^4 \\ &+ \frac{1}{2} e^{NN_0^2\beta_1} N \left( \frac{\left(\mathrm{erf}\left(\sqrt{\beta_1}\Lambda_4\right) - \mathrm{erf}\left(\sqrt{\beta_1}\Lambda_2\right)\right)\sqrt{\pi}}{\sqrt{\beta_1}} - 4e^{-\beta_1\Lambda_2^2}\Lambda_2 \right. \\ &\left. + 4e^{-\beta_1\Lambda_4^2}\Lambda_4 \right) N_0^2 - e^{-\beta_1\Lambda_1^2}\Lambda_1^3. \end{aligned} \tag{109}$$

$$F_{20} = e^{\beta_1(NN_0^2-\Lambda_2^2)}\Lambda_2^3 - e^{-\beta_1\Lambda_3^2}\Lambda_3^3 - e^{\beta_1(NN_0^2-\Lambda_4^2)}\Lambda_4^3. \tag{110}$$

$$F_{21} = \sqrt{\pi}\left(\mathrm{erf}\left(\sqrt{\beta_1}\Lambda_1\right) - e^{NN_0^2\beta_1}\mathrm{erf}\left(\sqrt{\beta_1}\Lambda_2\right) + \mathrm{erf}\left(\sqrt{\beta_1}\Lambda_3\right) + e^{NN_0^2\beta_1}\mathrm{erf}\left(\sqrt{\beta_1}\Lambda_4\right)\right). \tag{111}$$

$$\begin{aligned} F_{22} = &-e^{NN_0^2\beta_1} N \left(\mathrm{erf}\left(\sqrt{\beta_1}\Lambda_2\right) - \mathrm{erf}\left(\sqrt{\beta_1}\Lambda_4\right)\right)\sqrt{\pi}\sqrt{\beta_1}N_0^2 + e^{-\beta_1\Lambda_1^2}\Lambda_1 - e^{\beta_1(NN_0^2-\Lambda_2^2)}\Lambda_2 \\ &+ e^{-\beta_1\Lambda_3^2}\Lambda_3 + e^{\beta_1(NN_0^2-\Lambda_4^2)}\Lambda_4. \end{aligned} \tag{112}$$

$$\begin{aligned} F_{23} = &-e^{\beta_1(NN_0^2+\Lambda_1^2+\Lambda_2^2+\Lambda_3^2+\Lambda_4^2)} N \left(\mathrm{erf}\left(\sqrt{\beta_1}\Lambda_2\right) - \mathrm{erf}\left(\sqrt{\beta_1}\Lambda_4\right)\right)\sqrt{\pi}\sqrt{\beta_1}N_0^2 + e^{\beta_1(\Lambda_2^2+\Lambda_3^2+\Lambda_4^2)}\Lambda_1 \\ &- e^{\beta_1(NN_0^2+\Lambda_1^2+\Lambda_3^2+\Lambda_4^2)}\Lambda_2 + e^{\beta_1(\Lambda_1^2+\Lambda_2^2+\Lambda_4^2)}\Lambda_3 + e^{\beta_1(NN_0^2+\Lambda_1^2+\Lambda_2^2+\Lambda_3^2)}\Lambda_4. \end{aligned} \tag{113}$$

where,

$$N = Q_2(\delta(\alpha \mathrm{k} - \delta) + Q_3)^2, N_0 = \sqrt{-Q_2}, \Lambda_1 = \frac{N_0(N - \delta^2)}{2\delta}. \tag{114}$$

$$\Lambda_2 = \frac{N_0(\delta^2 + N)}{2\delta}, \Lambda_3 = \frac{N_0((\delta + \alpha \mathrm{k}\lambda)^2 - N)}{2(\delta + \alpha k\lambda)}, \Lambda_4 = \frac{N_0((\delta + \alpha \mathrm{k}\lambda)^2 + N)}{2(\delta + \alpha k\lambda)}. \tag{115}$$

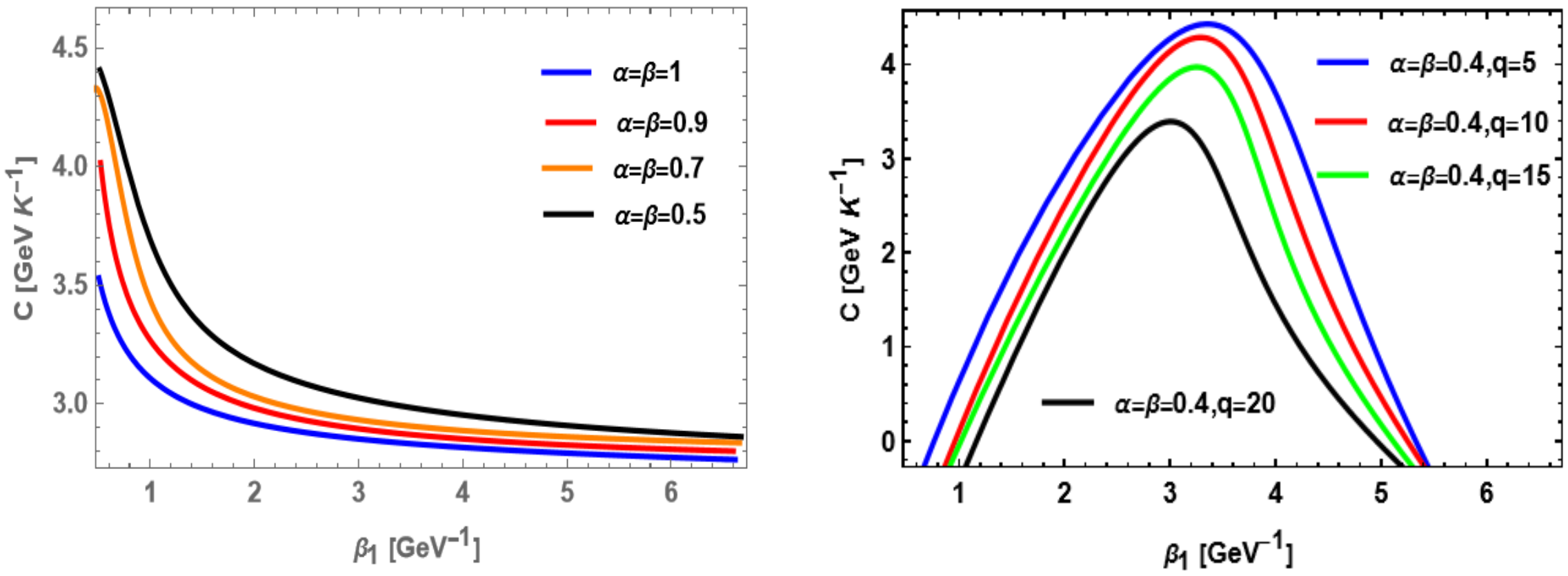

Figure 6: (The left panel), the specific heat (C) is displayed as a function of $\beta_1$ for various values of $\alpha$, and $\beta$. (The right panel), C is displayed as a function of $\beta_1$, for various values of deformed parameter q for CPSHEP at $\alpha = \beta = 0.4$.

# 5. Superstatistics Formulation

Superstatistics represents a statistical concept that applies to dynamic nonequilibrium systems and statistically intense variable ($\beta_1$) variation [50]. Chemical potential and energy fluctuation, which are primarily described in terms of the effectual Boltzmann variable, are included in this extensive parameter that experiences spatiotemporal volatility [51]. Apparently, Edet et.al [52], the actual Boltzmann factor is presented as follow.

$$B(E) = \int_0^\infty e^{-\beta_1' E} f(\beta_1', \beta_1) d\beta_1'. \tag{116}$$

where $f(\beta_1', \beta_1) = \delta(\beta_1 - \beta_1')$ is the Dirac delta function. Fortunately, the generalized Boltzmann constant is given as the following when stated in terms of the deformation parameter q:

$$B(E) = e^{-\beta_1 E}\left(1 + \frac{q}{2}\beta_1^2 E^2\right). \tag{117}$$

The partition function for superstatistics formalism is then provided as

$$Z_s = \int_0^\infty B(E) dn. \tag{118}$$

The modified Boltzmann constant equation is presented as when Eq. (90) is substituted into Eq. (117).

$$B(E) = \left[\frac{2}{2}\beta_1^2\left(\left(Q_1 + \frac{Q_2(\delta(k_1\alpha - \delta) + Q_3)}{2}\right) + \left(\frac{Q_2\rho^2}{4} + \frac{Q_2(\delta(k_1\alpha - \delta) + Q_3)^2}{4\rho^2}\right)\right)^2\right] * e^{-\beta_1\left[\left(Q_1+\frac{Q_2(\delta(k_1\alpha-\delta)+Q_3)}{2}\right)+\left(\frac{Q_2\rho^2}{4}+\frac{Q_2(\delta(k_1\alpha-\delta)+Q_3)^2}{4\rho^2}\right)\right]}. \tag{119}$$

Using Eq. (118), the superstatistics partition function equation is given as.

$$Z_s = e^{\beta_1\left(Q_1+\frac{Q_2(\delta(k_1\alpha-\delta)+Q_3)}{2}\right)} \int_0^\infty \left[1 + \frac{q}{2}\beta_1^2\left(\left(Q_1 + \frac{Q_2(\delta(k_1\alpha - \delta) + Q_3)}{2}\right) + \left(\frac{Q_2\rho^2}{4} + \frac{Q_2(\delta(k_1\alpha - \delta) + Q_3)^2}{4\rho^2}\right)\right)^2\right] * e^{\beta_1\left(\frac{Q_2\rho^2}{4}+\frac{Q_2(\delta(k_1\alpha-\delta)+Q_3)^2}{4\rho^2}\right)} d\rho. \tag{67}$$

The partition for superstatistics obtained from Eq. (120), is given as.

$$Z_s = \frac{\sqrt{\pi}\sqrt{\frac{1}{\beta_1}}\beta_1^{3/2} e^{-\frac{1}{2}\beta_1(NQ_2+NN_0^2+2Q_1)}\left(Q_2(\beta_1^2 q(NQ_2 + 2Q_1) + 4) - \beta_1 q Q_2\left(\sqrt{\frac{1}{\beta_1}}\beta_1^{3/2} NN_0^2 + 1\right)\right)}{4(\beta_1(-Q_2))^{3/2}} \tag{68}$$

where,

$$F_s(\beta_1) = -\frac{\ln(Z_s)}{\beta_1}. \tag{69}$$

$$U_s(\beta_1) = -\frac{\partial \ln(Z_s)}{\partial \beta_1}. \tag{70}$$

$$S_s(\beta_1) = K_1 \ln(Z_s) - \beta_1 K_1 \frac{\partial(\ln(Z_s))}{\partial \beta_1}. \tag{71}$$

$$C_s(\beta_1) = -K_1\beta_1^2 \frac{\partial U_s}{\partial \beta_1} \tag{72}$$

# 6. Results and Discussion

We plotted all figures using the following parameters $\beta_1 = \frac{1}{KT}$, and $K$ is the Boltzmann constant. When the principal quantum number n is around 0 and $1,2, \ldots, \lambda$, and $\mu$ is the reduced mass for quarkonium as $= \frac{m_1 * m_2}{m_1 + m_2}$. In Fig. 1 (left panel), the suggested coulomb plus screened hyperbolic exponential potential (CPSHEP) and an approximation one is plotted according to two characteristics: the coulomb potential and confinement potential characterize the short and long distances, respectively. The approximate potential is brief, with the actual potential reaching 0.8 fm. We plotted The Pekeris approximation graph compared to the screening parameter $\alpha_x$ as shown In Fig. 1(the right panel), since the approximation fits the anticipated potential. In Fig. (2), it shows how the temperature parameter affects the partition function for thermodynamic properties and superstatics. In this case, the partition function grows nonlinearly with $\beta_1$ whereas, the partition function in Fig. 2 (the left panel), deviated as increased, it later converged for the superstatistics, as seen in Figure. 2 (The left panel). As can be seen, the partition function (Z) is delicate at the highest values of $\beta_1$, the range of $\beta_1$ between $4\text{GeV}^{-1}$ and $7\text{GeV}^{-1}$ correlates with T between 0.142 and 0.25GeV. Additionally, it needs to be observed that as fractional parameters $\alpha$ and $\beta$ are increased, the curve's conduct becomes lower. In Fig. 2 (the right panel) (superstatics), it can also be seen that the partition function converging as acquires higher values of $\beta_1$, also by increasing deformation parameter q then curves become higher, deformation parameter q improves the behavior of superstatics. The behavior of $Z_s(\beta_1)$ is compatible with Ref. [53]. In Ref. [54], the ground state is shown as the node with the highest value on the probability density diagram. Temperature and the Z's behavior in relation to the largest possible quantum number($\lambda$). In addition, as the quantum number ($\lambda$) increases, Z decreases simultaneously increasing with temperature. According to Ref. [55], all diatomic molecules show a monotonic drop in Z as $\lambda$ increases. For some typical ranges of quantum number $\lambda$, Z achieves a constant value before declining further. Nevertheless, Z monotonically increases with increasing $\beta$. A closed-form representation of the temperature-dependent partition function $Z(\beta)$ was developed. In Ref. [70], the authors Plotted the vibrational partition function variation with a different of q values, it can be observed. that when $\beta$ rises, the partition function gets smaller. Furthermore, when q rises, Z decreases as $\beta$ increases. Furthermore, as q increases then Z does as well. In Fig. 3 (the left panel), the behavior of curves for free energy $F(\beta_1)$ increases monotonically by decreasing temperature ($\beta_1$), by increasing $\alpha$ and $\beta$ curves become higher using Eq. (95). This finding has a good agreement with Ref. [64]. In Fig. 3 (the right panel). For the superstatics, when the system's temperature ($\beta_1$) steadily drops, $F_s(\beta_1)$ rises monotonically. A decrease in the deformed parameter (q) always results in a larger free energy using Eq. (122). In Ref. [63], for CuLi and ScF, the free energy curves only decrease gradually with increasing $\beta$, but for various values of $\lambda$, the free energy curves of HCl and ScF increases and get lower with increasing $\beta$. In Ref. [70], when $\beta$ rises, the free energy falls monotonically. By increasing the deformation parameter q, the free energy curves rise and converge. As demonstrated in Fig. 4 (The left panel), and Fig. 4 (the right panel), we displayed the vibrational mean energy variation $U(\beta_1)$ versus superstatics and thermodynamic properties, respectively. Fig. 4 (the left panel), illustrates how $F(\beta_1)$ rises with decreasing values for all values of $\beta_1$. Additionally, curves rise when the values of the fractional parameters $\beta$ and $\alpha$ increase using Eq. (97). In Fig. 4 (the right panel), illustrates how the superstatics mean energy $U_s(\beta_1)$ changes when the system's temperature reduces. For different values of the deformed parameter (q), $U_s(\beta_1)$ decreases exponentially at a specific absolute temperature. According to Fig. 4(the right panel), the variation of curves converges for the superstatics. Additionally, the superstatics in Fig. 4(the right panel), show that $U_s(\beta_1)$ decrease as $\beta_1$ grows and curves become lower when q increases by using Eq. (123). This behavior is computable with Ref. [56], with rising $\beta$ and quantum number $\lambda, U_s(\beta_1)$ at first rises monotonically and subsequently reduces. Also, in Ref. [57], by increasing $\beta_s$ then U decreases. When the values of $\alpha$ and $\beta$ increase then the curves become lower as in Fig. 4(the right panel), $U_s(\beta_1)$ decreases with increasing $\beta$ and increases with increasing q as in Ref. [70]. In Fig. 5 (the left panel), we note that entropy $S(\beta_1)$ declines as the system's temperature rises ( $\beta_1$ dropping), and $S(\beta_1)$ associated with vibrations is diverging. In Fig. 5 (the right panel), the curves become lower when fractional parameters $\alpha$ and increase by using Eq. (102). In Fig. 5(the left panel), the superstatistics entropy $S_s(\beta_1)$ changes directly with temperature and inversely with $\beta_1$. This means that for any value of the deformed parameter, the system's disorderliness grows as the system's temperature rises. $S_s(\beta_1)$ converged when q increases then behavior of curves becomes higher, $S_s(\beta_1)$ has a turning point when $\beta_1$ equals 2 as shown in Fig. 5 (the left panel) using Eq. (124). This behavior is compatible with Ref. [53] and Ref. [58], which entropy (S) varies in reverse with respect to a range of values $\beta$. For light quark, odd quark, and natural particles, the authors of Ref. [59, 60, 61] showed that S increases with increasing temperature. In Ref. [70], the authors showed that as $\beta$ increases then the system's entropy decreases, the system's entropy rises when the deformation parameter q grows for a variety of values. In Fig. 6(the right panel), here is the heat capacity's behavior $C(\beta_1)$ against $\beta_1$ for various values of $\alpha$ and $\beta$. We note that $C(\beta_1)$ increases monotonically with decreasing $\beta_1$. Also, when $\alpha$ and $\beta$ decrease the curves get higher by using Eq. (108). In Fig. 6 (the left panel), for superstatistics, variations in the heat capacity $C_s(\beta_1)$ against $\beta_1$ is plotted for various values of the deformation parameter q, it is noticeable that $C_s(\beta_1)$ increases with decreasing $\beta_1$. Also, $C_s(\beta_1)$ converges when q increases, then the behavior of curves becomes lower as depicted. $C_s(\beta_1)$ has a turning point when $\beta_1$ equals 3 $\text{GeV}^{-1}$ using Eq. (125). This action has the same behavior for many works, such as in Ref. [62], for various values of $\lambda$, the author displayed specific capacity versus temperature $(\beta)$, C increases as $\beta$ decreases. As in Ref. [70], It is clearly visible that when q grows, the system's specific heat capacity falls while increasing monotonically with $(\beta)$. It's also significant to observe that normal statistics are returned when $q = 0$. In Ref. [56], the authors depict how specific heat $C(\beta)$ changes in relation to temperature. Specific heat grows monotonically as $\beta$ rises. and then declines as $\beta$ and $\lambda$ rise, whenever every plot reaches convergence. Each plot's convergence provides a measure of the variability in temperatures at which charmonium disintegrates into its component parts as quark charms.

# 7. Conclusion

We eliminated the parametric second-order differential equation with generalized fractional derivative and applying the NU method. Through the parametric generalized fractional Nikiforov-Uvarov (NU) approach, we arrive to the solution of the SE by employing Coulomb plus screened exponential hyperbolic potential. In addition, getting the special classical solution at $\alpha = \beta = 1$, then $k_1 = 1$ as in Ref. [49]. We estimated and plotted thermodynamic properties in the fractional form and discovered that our results are consistent with prior publications and that the physical behaviour of thermodynamic properties. The work was expanded to include thermal characteristics and superstatics, such as partition function $Z(\beta_1)$, free energy $F(\beta_1)$, mean energy $U(\beta_1)$, entropy $S(\beta_1)$, and heat capacity $C(\beta_1)$. Our results revealed that the effect of the generalized fractional derivative on the thermodynamic properties and superstatics are the same. We conclude that fractional parameters play an important role using CPSEHP potential. Also, fractional parameter has a good effect on the behaviour of thermodynamic and superstatistics curves. We also noticed that as free energy grows, internal energy, specific heat, and entropy decrease. Numerous studies, for example [67, 68, 69], do not explore the thermodynamic features and superstatics of heavy quarkonium in the fractional models. We conclude that the fractional parameter has a significant effect on thermodynamic properties.